\documentclass[twocolumn,showpacs,pra]{revtex4}
\usepackage{times}
\usepackage{graphicx}
\usepackage{graphics}
\usepackage{dcolumn}
\usepackage{amsmath}
\newcommand{\comment}[1]{}
\begin{document}
\def\mathbi#1{\textbf{\em #1}}
\title[]{Atom trapping with a thin magnetic film}

\author{Micah Boyd, Erik W. Streed, Patrick Medley, Gretchen K. Campbell, Jongchul Mun, Wolfgang Ketterle, David E. Pritchard}
\homepage[URL: ]{http://cua.mit.edu/ketterle_group/}
\affiliation{MIT-Harvard Center for
Ultracold Atoms, Research Laboratory of Electronics and Department
of Physics, Massachusetts Institute of Technology, Cambridge, MA
02139, USA}
\date{\today}
\pacs{03.75.Be, 03.75.Lm, 75.50.Ss, 75.70.i}

\begin{abstract}
We have created a $^{87}$Rb Bose-Einstein condensate in a magnetic trapping potential produced by a hard disk platter written with a periodic pattern.  Cold atoms were loaded from an optical dipole trap and then cooled to BEC on the surface with radiofrequency evaporation.  Fragmentation of the atomic cloud due to imperfections in the magnetic structure was observed at distances closer than 40 $\mu$m from the surface.  Attempts to use the disk as an atom mirror showed dispersive effects after reflection.
\end{abstract}

 \maketitle
Micrometer scale magnetic traps for Bose-Einstein condensates have been the focus of much experimental work since their first demonstration \cite{Ott01, Hansel01}.  Their potential uses in atom interferometry, precision measurements, and experiment miniaturization have motivated many groups to develop sophisticated techniques for manufacturing and controlling these atom chips \cite{Shin05,Hall05,Sinclair05}.  However, since the earliest inceptions of atom chips, physical imperfections in the chip surface have lead to perturbations in the trapping potential that prohibit the coherent manipulation of atoms close to the surface\cite{LeanhardtSurface02,Lin04}.  Permanent magnets offer a possible solution to several problems inherent to current carrying wire traps.  First, the magnets are almost completely decoupled from the rest of the laboratory, minimizing the effects of environmental electrical noise.  Second, they do not require current to be sourced or sinked, enabling designs that would be too complicated or impossible to create using electromagnets.  Last, extremely high field gradients are possible close to magnetic domain boundaries, whereas traditional atom chip operation is limited by heat dissipation from small wires.  Although a substrate of magnetizable material may be made extremely smooth and uniform, imperfections in the process of etching \cite{Lev03} or writing magnetic structures could lead to some of the same problems as with wire-based designs.  It is not yet known which technique will provide the best performance.

In this paper we investigate atom trapping with a thin magnetic film; specifically that of a hard disk platter written with a periodic pattern.  This approach offers some advantages over previous work on neutral atom trapping and BEC creation using permanent magnets \cite{Sinclair05,Hall05,Wu06}.  Here we use a thin metallic film with a large remnant magnetization from a commercial product which has already been refined to a high degree and a writing technique more accurate than anything previously demonstrated.  Cold $^{87}$Rb atoms were first loaded into the magnetic potential formed by the disk and used RF evaporation to produce BEC.  By changing the trapping potential, the atoms could be pushed closer to the surface of the disk to probe for imperfections in the potential.  Finally, the BEC was dropped onto the disk from a height of 2.7 mm in an attempt to produce a specular reflection of the atomic cloud.

The form of the written magnetization on the disk is of critical importance.  A surface magnetization of the form
\begin{equation}
\label{eq:surfacemag}
\mathbi{M(x)}=M_{0}\cos{\left(kx\right)}\hat{y},
\end{equation}
where $M_0$ is the magnetization of the material and $k$ is the wavevector of the sinusoidal pattern produces a magnetic field above the surface
\begin{equation}
\left(B_{x},B_{y},B_{z}\right)=B_{0}e^{-ky}\left(-\cos{\left(kx\right)},sin{\left(kx\right)},0\right),
\end{equation}
which is of uniform coplanar magnitude and decays exponentially away from the surface \cite{Hinds99}.  This is an ideal potential for reflecting weak field seeking atoms.

\begin{figure}
\centering{
\includegraphics[hiresbb=true,width=0.45\textwidth]{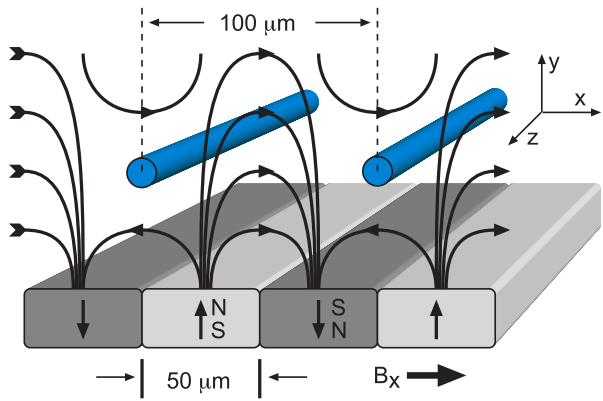}
\caption{\label{fig:surface_b_diagram} Geometry of the magnetic trap formed by a magnetic film.  Magnetic field lines resulting from the addition of a radial ($\hat{x}$) bias field to the magnetized surface.  The tubes represent the locations of the field minima where atoms are trapped.  The addition of an axial bias field of about 1.0 G along $\hat{z}$prevents atoms loss from Majorana spin flips.}
}
\end{figure}

The addition of a bias field $B_x$ along $\hat{x}$ (or $B_y$ along $\hat{y}$) produces a series of quadrupole shaped field minima above alternating tracks, depicted in Fig. \ref{fig:surface_b_diagram}.  A second bias field $B_{z}$ along $\hat{z}$ removes the magnetic field zero and weak field seeking atoms may be trapped near the surface with radial trap frequency
\begin{equation}
2\omega_r = kB_x\sqrt{\frac{\mu_{B}g_{F}m_{F}}{mB_{z}}}
\end{equation}
where $m$ is the atomic mass, and $\mu_{B}g_{F}m_{F}$ is the Zeeman energy \cite{Hinds99, Sinclair05}.

The magnetic media used in this work was a prototype hard disk with a radius of 65 mm and a thickness of 0.635 mm.  The magnetizable material is a dual layer system with a 20 nm thick magnetic Co-Cr-Pt based oxide layer and a 200 nm thick ``magnetically soft underlayer'', both of which are also good conductors.  The substrate is glass, the coercivity is 6.8 kOe and the magnetic remnant is 0.64 memu/cm$^2$.  The easy axis of magnetization of this prototype ``out-of-plane'' disk is aligned normal to the surface of the disk, as opposed to most modern commercial hard disks which are magnetized in the plane of the surface.  This alignment of the magnetic domains produces about 6000 G at the surface.

\begin{figure}
\centering{
\includegraphics[hiresbb=true, width=0.49\textwidth]{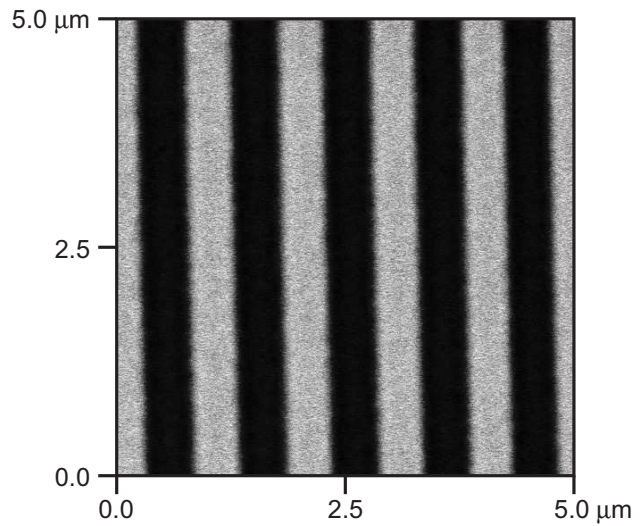}
\caption{\label{fig:mfm_image} Magnetic force microscopy image of the disk used in this experiment.  This scan is over the region with $\lambda$ = 1.0 $\mu$m.  The track edges are of the same uniformity in the  $\lambda$ = 100.0 $\mu$m region.  The characteristic size of the roughness of the domain boundaries is about 30 nm.} 
}
\end{figure}

The disk was provided to us pre-written using a Guzik spin stand, which is essentially a hard disk read/write head with absolute positioning capability.   This method of writing allows for the creation of truly arbitrary pattern impossible to create with larger scale magnetic writing devices \cite{Sinclair05} or physical structures \cite{Hall05}, and produces smaller and cleaner structures than optical writing techniques \cite{Eriksson04}.  Two patterns were used in this experiment, written on different radial regions of the same disk.  The first region had alternating stripes of up and down magnetization with a period of $2\pi/k=\lambda$ = 100.0 $\mu$m and the second had $\lambda$ = 1.0 $\mu$m.  The media has a very square hysteresis loop which precludes a purely sinusoidal magnetization.  The pattern is instead written as a square wave, and the higher Fourier harmonics should be negligible at the height where we trap atoms \cite{Hinds99}.  Figure \ref{fig:mfm_image} shows a magnetic force microscopy image of the disk after writing which shows a significantly less noisy pattern than magneto-optical thin films written with laser beams \cite{Lau99}.  Atomic force microscopy showed a physical roughness of $\sim$3 nm over a region of 50 $\mu$m. 

In our experiment, cold atoms were delivered to the surface in a two step process.  First, cold $^{87}$Rb atoms in the $F=1,m_f=-1$ state were created in a Ioffe-Pritchard magnetic trap with $T\gtrsim T_c$ where $T_c$ is the critical temperature for Bose-Einstein condensation.  The atoms were then transferred to an optical dipole trap and transported 36 cm to a separate auxiliary chamber (see \cite{Streed06} for more details).  The optical dipole trap was formed by a $\lambda$ = 1064 nm laser focused to a 30 $\mu$m $1/e^2$ radius spot, and the focus (with the atoms) was translated into the auxiliary chamber as described in \cite{Gustavson02}.  By transporting atoms just above $T_c$, the cloud was less sensitive to vibrations, and higher laser powers could be used without causing rapid three-body losses.

The atoms were loaded onto the surface from the optical trap by translating the focus of the optical trap to a position parallel to and 50 $\mu$m above the region with $\lambda$ = 100~$\mu$m.  A Z-shaped wire below the disk provided axial confinement, and a small $B_x$ created radial trapping on the surface.  The optical trap was ramped off over 2 seconds, transferring the atoms with almost unity efficiency.  The optical trap enabled loading of $>$90\% of the atoms into a single surface trap site.  RF Evaporation over 20 seconds from 1.200 MHz to 0.890 MHz produced a BEC with approximately 50,000 atoms in a trap with $(\omega_x,\omega_y,\omega_z)=(390, 390, 9)\times2\pi$ Hz.  The spatial distribution was clearly bimbodal, and condensate fractions of $>$80\% were observed.

\begin{figure}
\centering{
\includegraphics[hiresbb=true, width=0.45\textwidth]{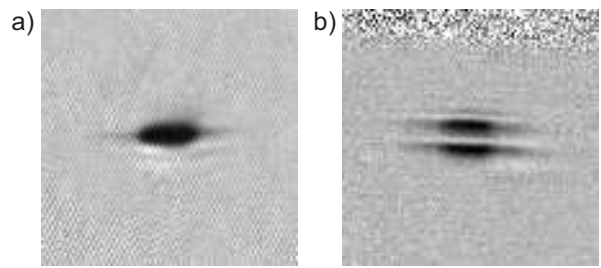}
\caption{\label{fig:bec_images} Absorption image of the BEC in trap near the surface.  a) Top imaging.  The field of view is 0.58 mm $\times$ 0.58 mm.  b) Side imaging.  The field of view is 0.56 mm $\times$ 0.56 mm.  The double image comes from the reflection of the imaging beam in the platter surface.  The trap frequencies were $2\pi\times$(400, 400, 10) Hz.  The configuration of the experiment prevented ballistic expansion, but the bimodal distribution in trap is still clear.}
}
\end{figure}

The atomic cloud was detected with on-resonance absorption imaging (Fig. \ref{fig:bec_images}).  The platter is a good reflector ($>$95\%) for 780 nm light, so grazing incidence imaging was used to measure the distance from the atoms to the surface.  Normal incidence imaging was also used, in which case the imaging light passed through the atoms twice, bouncing off the disk.  The lack of physical structures on the surface resulted in good image quality.

\begin{figure}
\centering{
\includegraphics[hiresbb=true, width=0.45\textwidth]{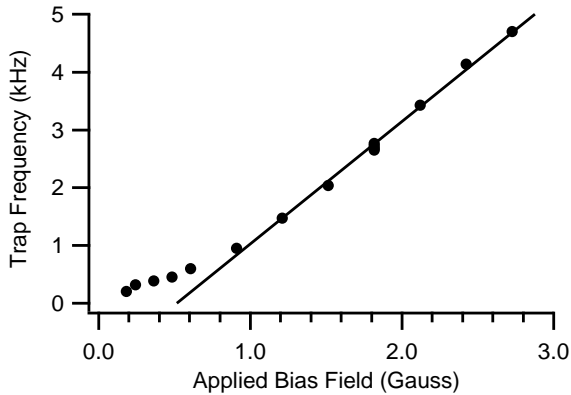}
\caption{\label{fig:trapfrequencies} Radial trap frequency vs applied radial magnetic field.  Atoms were loaded into a single trap site on the surface and evaporated to BEC.  The trap frequencies were measured with parametric heating, but above 5 kHz other heating effects made it difficult to resolve reliably.  The deviation from linearity at low applied fields is most likely due to a small, off-axis, residual bias field.}
}
\end{figure}

The axial trap frequency in the surface trap was measured by imaging oscillations of the atomic cloud, and radial trap frequencies were measured by parametric heating [Fig. \ref{fig:trapfrequencies}].  Attempts to release the atoms from the trap and observe ballistic expansion in time of flight were hindered by the geometry of the system and the static nature of the magnetic surface, so all of the imaging was done in-trap.  The Z-wire trap was left on at all times to provide axial trapping, but its effect on the radial trapping was negligible.  Ramping up the current in external electromagnets increased $B_x$, and $\omega_r$ as high as $2\pi\times5$ kHz was measured.  Using a disk with $\lambda$ = 10.0 $\mu$m, we have measured $\omega_r$ as high as $2\pi\times16$ kHz.  High transverse trap frequencies are desired for studies of 1D systems \cite{Kim05}, but atom heating and loss at higher trap frequencies prevented such studies here.

While trapped 40 $\mu$m above the surface, the BEC had a lifetime of $\sim$30 s.  Previous experiments with atomic clouds magnetically trapped near conducting surfaces showed atom loss resulting from spin flips driven by Johnson noise \cite{Jones03, Lin04, Scheel05}.  For this experiment however, the conducting layer was extremely thin (200 nm), and the atoms were relatively far from the surface.  These two parameters reduced any spin flip loss below the level of our 1-body losses at a vacuum pressure of $\sim5\times10^{-11}$ Torr.

\begin{figure}
\centering{
\includegraphics[hiresbb=true, width=0.45\textwidth]{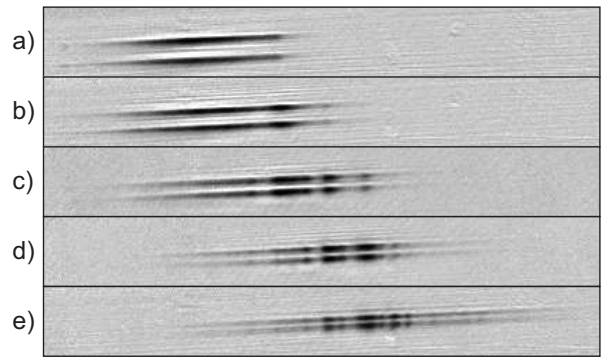}
\caption{\label{fig:breakup} Breakup of the atomic cloud as the atoms approached the surface.  The thermal cloud was evaporated to $T\approx T_c$, then the axial trapping frequency was reduced to from 10 Hz to ~1 Hz to allow the atoms to expand in one dimension.  The axial bias field was then increased to push the atoms closer to the surface, and the atoms were imaged in trap.  The grazing incidence imaging used here produces a double image of the atoms from the primary and secondary reflections of the imaging beam off the disk.  The field of view is 0.29 mm $\times$ 2.3 mm, and each frame is an average of five distinct images to reduce noise for clarity.  At distances smaller than 40 $\mu$m, significant breakup was observed.  Height for the images shown: a) 35 $\mu$m b) 29 $\mu$m c) 25 $\mu$m d) 22 $\mu$m e) 19 $\mu$m.}
}
\end{figure}

The BEC was also used to probe imperfections in the trapping potential.  After evaporation, the axial trapping frequency was reduced from 10 Hz to $\sim$1 Hz to allow the cloud to expand slightly in one dimension to increase the sensitivity and the measurement area, and the radial bias field ($B_r$) was increased to push the atoms closer to the surface.  At distances closer than 40 $\mu$m, breakup of the atomic cloud was observed [Fig. \ref{fig:breakup}], and the strength and spatial frequency of the perturbations increased as the atoms neared the surface, similar to \cite{LeanhardtSurface02, Cassettari00}.  The magnitude and size scale of these imperfections can be attributed to the sputtering process used to create the film.  Magnetic grain nucleation occurs randomly, resulting in small angular misalignments of the axes of anisotropy of individual magneto-crystalline domains (typically a few degrees), the individual magnetic grains can vary in size by about 25\% of their $\sim$7 nm diameter, and the magnetic moments have a distribution of magnitudes.  All of these phenomena create imperfections in atomic trapping potentials.

While the presence of these perturbations does not necessarily preclude the creation of neutral atom  waveguides using this type of substrate, it does limit its usefulness.  High trapping frequencies and single transverse mode confinement require close proximity to the surface and will suffer from the observed imperfections.  However, for use as a high reliability, low noise, and low cost waveguide at distances $>$50 $\mu$m, commercial metallic, magnetic thin films are ideal.

The magnetized surface was also examined for its usefulness as an atom mirror, similarly to \cite{Roach95, Sidorov96, Lev03}.  In order to minimize residual and time dependent magnetic fields (from magnetic trap turnoff), all optical trapping was first used to create a BEC.  The optical dipole trap provided only weak axial confinement, so a second, 200 mW laser beam with $\lambda$ = 1064 nm and 200 $\mu$m waist was added in a crossed configuration, and allowed more efficient evaporation.  To evaporate to BEC the power was reduced over 2.0 s.  An axial bias field $B_z$ = 1.2 G was maintained throughout the experiment, and $B_x$ and $B_y$ were minimized to $<$20 mG.

The region of the disk used as a magnetic mirror was written with $\lambda$ = 1.0 $\mu$m.  This value was chosen to be large compared to the magnetic writing precision, but small compared to the extent of the atomic cloud.  The BEC was released from the crossed ODT 2.7 mm above the disk.  The atoms fell under gravity for 23.5 ms and then reflected off the magnetic potential.  The atomic cloud was imaged (in separate runs) from both the top and side after various times of flight.  Side imaging showed that the reflection from the disk did not significantly effect the axial or vertical velocity distributions of the atomic cloud.  Top imaging, however, showed significant spreading along the vector of the magnetic pattern after reflection, analogous to bouncing off a rough mirror \cite{Perrin05}.  The magnitude of this dispersion was minimized by fine tuning $B_x$ and $B_y$, but was impossible to eliminate in our apparatus.  Before bouncing off the disk, the \^x width of the cloud expanded by 1 mm/s, and after bouncing the \^x width increased by 34 mm/s.  The reflection was not performed on the region with  $\lambda$ = 100.0 $\mu$m, but experiments using a different disk with $\lambda$ = 10.0 $\mu$m showed similar effects.

The magnetic properties of the hard disk surface are one possible source of the observed expansion after reflection.  While the theory presented earlier (Eq. \ref{eq:surfacemag}) applies to a sinusoidally magnetized surface, the square hysteresis loop and directional anisotropy of digital recording media prevent recording of a pure sine wave, instead forming a square wave approximation of that sine, as discussed in \cite{Hinds99}.  While the higher harmonics resulting from this approximation do not necessarily adversely effect BEC trapping, they create a corrugation of the planar equipotential that prevented the specular reflection of a macroscopic atomic cloud.  Another contribution to the roughness of the reflection potential is the existence of small, stray magnetic fields.  Any nonzero component of $B$ in the $\hat{x}\hat{y}$ plane creates a regular corrugation in the plane, further inhibiting specular reflection.

In conclusion, we have demonstrated BEC production on the surface of a hard disk platter.  Small scale imperfections in the magnetization caused condensate fragmentation close to the surface, prohibiting its use as a neutral atom waveguide.  The disk was also used as an atom mirror, and specular reflection was observed on two axes.  These results are a substantial improvement over that for early wire based experiments, and the possibility for more complex structures enables trap geometries impossible for electromagnets.  Microtraps based on permanent magnets, and in particular magnetic metallic thin films, may become an alternative to atom chips using current-carrying wires if the fabrication can be further improved, e.g. by using molecular beam epitaxy, and the writing process improved by using write heads optimized for recording DC structures on perpendicular media.

The authors would like to thank Aaron Leanhardt for experimental assistance, Tom Pasquini for a critical reading of the manuscript, and Min Xiao at Hitachi Global Storage Technologies for writing several disks for us and providing the MFM images, this work would not have been undertaken without her assistance.

\bibliography{platter}

\end{document}